\documentclass[3p,times,twocolumn]{elsarticle}
 \biboptions{comma,sort&compress}
\usepackage[utf8]{inputenc} 
\usepackage{graphicx}
\usepackage{amsmath,amssymb,latexsym,amsbsy,amsmath,bm}
\usepackage{here}
\usepackage{soul,xcolor}
\usepackage{ecrc}

\def\vec#1{\bm{#1}}
\volume{00}

\firstpage{1}

\journalname{Nuclear and Particle Physics Proceedings}
\runauth{Jean-Marc Richard, Alfredo Valcarce, Javier Vijande}

\jid{nppp}
\jnltitlelogo{Nuclear and Particle Physics Proceedings}

\usepackage{amssymb}

\usepackage[figuresright]{rotating}
\usepackage{aas_macros}

\begin{document}

\begin{frontmatter}

\title{
%
Doubly-heavy tetraquark bound states and resonances\,$^*$} 
 
 \cortext[cor0]{Invited talk presented  by J.-M. Richard at QCD22, Montpellier (France), July 4-7, 2022. }

 \author[label1]{Jean-Marc Richard
 \corref{cor1} 
 }
   \address[label1]{Institut de Physique des 2 Infinis de Lyon\\
   Université Claude Bernard, CNRS-IN2P3-Universi\'e de Lyon\\
   4 rue Enrico Fermi, 69622 Villeurbanne, France
}
\author[label2]{Alfredo Valcarce
 \corref{cor2} 
 }
   \address[label2]{Departamento de F\'\i sica Fundamental,\\ 
Universidad de Salamanca, E-37008 Salamanca, Spain
}
\author[label3]{Javier Vijande
 \corref{cor3} 
 }
   \address[label3]{ Unidad Mixta de Investigacin en Radiofísica e Instrumentación Nuclear en Medicina (IRIMED)\\
   Instituto de Investigación Sanitaria La Fe (IIS-La Fe),Universitat de Valencia (UV) and IFIC (UV-CSIC)
Valencia, Spain}
\pagestyle{myheadings}
\markright{ }
\begin{abstract}
\noindent
We review the predictions of the quark model for the doubly-heavy tetraquarks $QQ\bar q\bar q$. 
The possibility of resonances near the $BB^*$ threshold in addition to a deeply bound state
is discussed.  
 
\end{abstract}
\end{frontmatter}
\section{Introduction}
In July 2021, the LHCb collaboration announced the discovery of a very sharp peak in the $DD\pi$ spectrum that was dubbed $T_{cc}$ \cite{LHCb:2021vvq,LHCb:2021auc}. It corresponds to a minimal quark content $cc\bar u\bar d$. 

The possibility of bound states in configurations $QQ\bar u\bar d$, due to the chromoelectric interaction, was predicted 40 years earlier~\cite{Ader:1981db}, and followed by several further studies. See, e.g., \cite{Heller:1985cb,Zouzou:1986qh,Carlson:1988hh,Semay:1994ht,Janc:2004qn,Vijande:2009kj,Richard:2018yrm} and 
references therein.

In the late 70s, the properties of chromoelectricity and in particular the consequences of flavor independence were already stressed, e.g., to explain the larger binding of the bottomonium with respect to its Zweig-allowed threshold as compared to charmonium. The role of chromoelectricity was not yet investigated in the multiquark sector, that was the realm of chromomagnetism~\cite{Jaffe:2004ph}. 

In the early 80s, it was shown that $QQ\bar q\bar q$ becomes bound in a given spin-independent central potential if the mass ratio $M_Q/m_q$ is large enough. It was later realized that the mechanism that stabilizes $QQ\bar q\bar q$ at large $M_Q/m_q$ is the same that makes the hydrogen molecule much more stable than the positronium one in atomic physics \cite{Richard:1992cb,Richard:1993zx}. 

It was also soon acknowledged that a pure chromoelectric binding of $QQ\bar q\bar q$ requires rather elusive quark mass ratios $M_Q/m_q$. Thus, for realistic situations such as double charm or double beauty, one needs the help of chromomagnetism, which is attractive if $\bar q\bar q=\bar u\bar d$ is in a color $\bar 3$ spin-isospin singlet state \cite{Janc:2004qn,Lee:2009rt}.

Along the years, the stability of doubly-heavy tetraquarks was confirmed or rediscovered in several approaches that are more ambitious than simple quark models. See, for instance, the talk by Marina Nielsen at this Conference~\cite{Abreu:2022vmj}. 

Interestingly, the $T_{cc}^+$ misses binding by a very small amount. So, it is almost certain that the heavier partners, $T_{bc}=\bar u\bar d$ and $T_{bb}=bb\bar u\bar d$ will be stable against dissociation. This will open a new chapter of weak interaction. There are already estimates of the lifetime, pointing out some of the leading decay channels. In particular, a lifetime much larger than for $bbq$ baryons could help to distinguish $T_{bb}$ in the final states at high energy colliders. There are at least two reasons for this large lifetime of $T_{bb}$, as compared to ordinary mesons and baryons carrying beauty: a fraction of the decay energy is consumed by the recoil of the spectator quarks; after, say, $b\to c+ \cdots$, the color and spatial overlap with the charmed particles of the final state is suppressed. 

In this contribution, we shall discuss the tetraquark spectrum as predicted in potential models, and in particular around the $BB^*$ threshold. The possibility of a second excited state and/or of resonances has been suggested. 
\section{Model}
Some benchmark calculations of tetraquarks have been done with the potentials built in \cite{Semay:1994ht}, which have been adjusted to fit ordinary hadrons. They read%
\begin{align}\label{eq:models}
 &V=\sum_{i<j} \tilde\lambda_i.\tilde\lambda_j \left[u(r_{ij})+\vec{\sigma}_i.\vec{\sigma}_j\,w_{ij}(r_{ij})\right]~,\notag\\[-3pt]
&u(r)=-\frac{\kappa}{r}+\lambda\,r^p-\Lambda~,\\[-3pt]
&w_{ij}(r)=\frac{2\,\pi\,\kappa'}{3\,m_i\,m_j}\,
\frac{\exp\left(-r^2/a_{ij}^2\right)}{\pi^{3/2}\,a_{ij}^3}~,
\quad a_{ij}=A\,\genfrac{(}{)}{}{}{2\,m_i\,m_j}{m_i+m_j}^{-B}~,\notag
\end{align}
where $m_i$ are the quark masses, $\tilde\lambda$ and $\vec{\sigma}$ the color and spin operators, respectively. 
The choice $p=1$ (AL1) of a linear confinement has been adopted, e.g., in \cite{Semay:1994ht} 
{and in \cite{Janc:2004qn} with a revised set of parameters},
while the choice $p=2/3$ (AP1) has been recently used by Meng et al.\ \cite{Meng:2020knc} 
with a revised set of parameters.
For estimating the binding energies of tetraquarks, we use consistently the threshold masses  computed in the same model, which are anyhow very close to the experimental ones. Note the smearing of the hyperfine interaction, with a range that depends on the masses. 

The four-body problem is solved for the mass configurations $MMmm$, with $m=m_d=m_u$ in the isospin limit, and $M=m_c$ or $m_b$. We use a Gaussian expansion
\begin{equation}\label{eq:Gauss}
 \Psi=\sum_{k,j} \sum_{n=1}^N \gamma^{k,j}_n \,\left[\exp\left(-\tilde X.A^{k,j}_n.X/2\right)+\cdots\right]~,
\end{equation}
where $X$ is a set of three Jacobi variables for the internal motion, the $A_n$ define-positive symmetric matrices, and the ellipsis corresponds to terms deduced by symmetries. The linear parameters $\gamma_n$ are determined by diagonalization, and the matrices $A_n$ by suitable minimization. Note that we use a single set of coordinates, but  matrices $A_n$  that are not necessarily diagonal, at variance with \cite{Meng:2020knc}, but the two procedures are equivalent. Anyhow, this is a very delicate four-body calculation, that requires a subtle mixture of long- and short-range terms, and a full account of all allowed color-spin configurations. 
\section{Bound states}
We now present our preliminary results. More details will be given elsewhere \cite{RVV}. We use  first the models AP1 with the parameters of \cite{Meng:2020knc} and  the original AL1 model \cite{Semay:1994ht}. As we shall see, both models tend to overbind $cc\bar u\bar d$, AP1 due to its smooth confinement \cite{Richard:2018yrm}, and AL1 by its chromomagnetic interaction that is slightly too attractive in the heavy-light sector. As an educated guess, before performing a more detailed tuning \cite{RVV}, we repeated the calculation using a combination, dubbed AL1', which consists of the central part of AL1 and the spin-spin part of AP1. It also reproduces the masses of the ordinary hadrons entering the thresholds. This new model is almost satisfactory for $cc\bar u\bar d$, as it gives a very tiny (but still slightly too large) binding below the nominal $DD^*$ threshold.

{Due to the 
absence of antiymmetrization for the heavy quarks, the $bc\bar u\bar d$ system has a lower bound state
with $J^P=0^+$, not allowed for tetraquarks with identical heavy flavors. Thus, the $J^P=1^+$ $bc\bar u\bar d$
state would decay electromagnetically to $\bar B D \gamma$ \cite{Carames:2018tpe,Francis:2018jyb}.}

\begin{table}
 \caption{Binding energies (MeV) of the lowest $J^P=1^+$ $QQ\bar u\bar d$ tetraquarks using some potential models. 
For AP1, more precise values are given in \cite{Meng:2020knc}.}
 \begin{center}
 \begin{tabular}{lcccc}
 \hline\\[-.35cm]
 Model& $cc\bar u\bar d$ & $bc\bar u\bar d$ & $bb\bar u\bar d$ & $bb\bar u\bar d^*$ \\[.1cm] 
 \hline\\[-.4cm]
  AP1 & 21 & 37 & 171 & 2.9  \\
  AL1 & 13 & 23 & 151 & 0.6  \\
  AL1'&  \phantom{0}8  &  18  & 133 & -     \\
\hline\\[-1.1cm]  
 \end{tabular}
 \end{center}
\label{tab:bs}
\end{table}

\begin{figure}[ht!]
 \centering
 \includegraphics[width=.8\columnwidth]{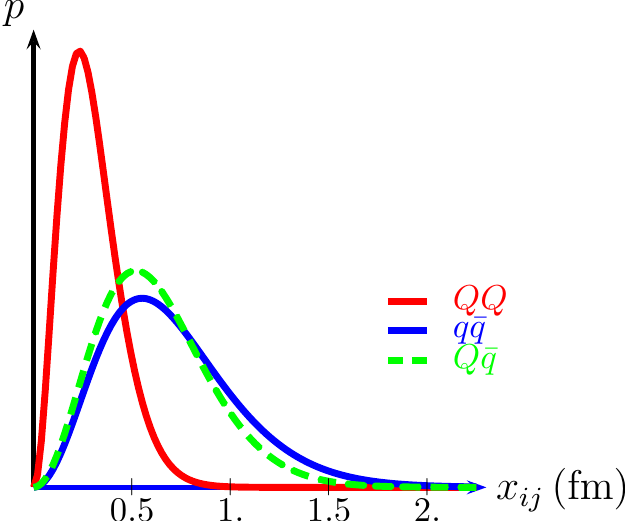}
 \caption{Radial distribution of the $QQ$, $Q\bar q$ and $q\bar q$ separations for the $bb\bar u\bar d$ bound state within the AP1 potential. }
 \label{fig:fig1}
\end{figure}

Some radial distributions $p(r_{ij})$ 
{for the $bb \bar u \bar d$ ground state} 
are shown in Fig.~\ref{fig:fig1}. The r.m.s.\ are clearly in order $\bar r(QQ)<\bar r(Q\bar q)\lesssim\bar r(q\bar q)$. The analogs for the  excited state in the same model AP1 are shown in \cite{Meng:2020knc}. 

The occurrence of a second bound state in the model AP1 was  pointed out by Meng et al.\ \cite{Meng:2020knc}. Such a state would decay radiatively towards the ground state or to a $BB$ pair that would decay weakly. To our surprise, this radial excitation survives in the less attractive AL1 model, with a much smaller binding, but it disappears in the more realistic AL1' variant. 

The above results have been obtained from straightforward four-body calculations. Of course, any approximation is welcome, if it simplifies the estimates and sheds some light on the inner structure.  The diquark-antidiquark scheme has been adopted by several authors, either as a phenomenological model or a way to escape the tedious four-body problem. For sure, there is a pronounced $QQ$ clustering in doubly-heavy tetraquarks, but the $\bar q\bar q$ pair remains widely spread. Moreover, a crude diquark approximation tends to overbind \cite{Richard:2018yrm}, with the risk of predicting stable states that are not there in a given model. 

The Born-Oppenheimer approximation \cite{1927AnP...389..457B} is one of the most important tool of quantum chemistry, and in the few-body community, the following empirical rule is transmitted  between the generations: \textsl{The Born-Oppenheimer works always better than expected}. It was applied to double-charm baryons, charmonium (the gluon field playing the role of the electrons in molecular physics), hybrids and tetraquarks. See, e.g., \cite{Braaten:2014qka} and refs.\ there. A comparison of the effective $QQ$ potentials for $QQ\bar q\bar q$ and $QQq$ is rather instructive \cite{Richard:2018yrm}. The two curves are almost identical, up to an additive constant. This provides a microscopic derivation of the mass relation \cite{Eichten:2017ffp}
\begin{equation}
 QQ\bar q\bar q= QQq+ Qqq-Q\bar q~.
\end{equation}
Note that this is at variance with atomic physics: the effective $pp$ potentials are rather different for $\mathrm{H}_2{}^+$ and $\mathrm{H}_2$: unlike the color in the QCD case, the charge of $pp$ is not neutralized in the 3-body configuration.  
The Born-Oppenheimer approximation to tetraquarks has also been studied recently by the Rome group \cite{Maiani:2022qze}, but without comparison with the doubly-heavy baryons.

\section{Resonances near threshold}
The calculation of bound states is already rather delicate in the  models such as \eqref{eq:models}, but the search for resonances is even harder. A horrible solution consists of using a simple trial function (for instance small $N$ in \eqref{eq:Gauss} or similar), and identifying the variational energy $E$ above the threshold as the energy of the resonance. Indeed, refining the wave function lowers the energy $E$ which would eventually coincide with the threshold. 

Several methods have been elaborated and successfully tried on toy models or on electron resonances in atomic physics. See, e.g., \cite{1998PhR...302..212M,2013JPhA...46h5303K,2018JPhB...51x5001S}. In the quark model, the method of real scaling (or stabilization) has been used in the case of the pentaquark~\cite{Hiyama:2018ukv}. With a model similar to \eqref{eq:models}, there is clear separation between genuine resonances and states that simply mimic the continuum. The idea is that the expectation value of an Hamiltonian is stationary near a resonance. In practice, the expectation value is calculated with a wave function of type \eqref{eq:Gauss}, in which some range parameters are varied, the ones corresponding to a separation into two hadrons. See, e.g., \cite{Meng:2021yjr}.  One notices, however, that the calculation is rather sensitive to the choice of the trial function on which real scaling operates. Thus, it is desirable to check the results against any change of the parameters, and to use alternative methods such as complex scaling, external confinement whose radius is varied, etc. 
\section{Weak decay of heavy tetraquarks}
The field of $\beta$-decay has been dramatically enlarged by the discovery of strangeness \cite{Leprince:1944}, later by the discovery of open charm at SLAC \cite{Goldhaber:1976xn}, and also by the discovery of open beauty at Cornell \cite{CLEO:1980oyr} (in the two latter cases, the $Q\bar Q$ quarkonium was found earlier). For an introduction, see, e.g., \cite{Pais:107749,Ezhela:317696,richard2021introduction}. Several intriguing properties have been found, such as the suppression of semileptonic decay of the $\Lambda$ hyperon, the spread of lifetime of the particles carrying charm, etc. 

One expects interesting developments on the decay mechanisms from hadrons with two heavy quarks, such as $b\bar c$ or $bcu$. For the tetraquark $bb\bar u\bar d$, there are already estimates of the lifetime and dominant modes, but with rather different predictions \cite{Xing:2018bqt,Ali:2018xfq,Hernandez:2019eox}. For instance, the lifetime is sometimes estimated to be about the same as that of $B$ mesons or $\Lambda_b$ baryon, or even slightly smaller. In contrast, the study in \cite{Hernandez:2019eox} predicts a significantly larger lifetime. The reason is that the color configuration and the spatial distribution in the final state do not match very well the properties of the final state hadrons, contrary to what happens for $B$ or $\Lambda_b$ decay.  An even simpler reason is that the released energy of a typical mode $b\to c+X$ benefits entirely to $B\to D+X$ in the case of a meson decay, while it is shared with spectators in the case of tetraquark decay; and it is notorious that weak interactions are very sensitive to the available phase-space. 
\section{Conclusions}
The long awaited discovery of the $T_{cc}^+$ opens a new era in the field of exotic hadrons, with compact states that probe new color configurations of the quark dynamics and multiquark states that decay weakly. The $T_{cc}^+$ is just at the edge between resonances, sometimes described as hadron-hadron molecules, and  compact bound states that probe the quark interaction. The binding of doubly heavy tetraquarks results from a conspiracy of chromomagnetic forces in the light sector, and chromoelectric effects in the heavy sector. 

The question has already been raised of what happens in the $bb$ sector: just a deeply bound $bb\bar u\bar d$ with $J^P=1^+$ or, in addition, other bound or resonant states very close to the threshold?
\vskip -.2cm
\paragraph{Acknowledgments:} We thank the organizers of QCD22 for maintaining the tradition of this conference in a stimulating and relaxing atmosphere. 
%

\end{document}